\documentstyle[aps,psfig,multicol]{revtex}
\begin{document}

%%
%%  TITLE PAGE
%%
\draft

\title{The Anisotropic Bak--Sneppen Model}

\author{D A Head\footnote{Electronic address: David.Head@brunel.ac.uk}}

\address{Institute of Physical and Environmental Sciences,
Brunel University, Uxbridge, Middlesex, UB8~3PH, United Kingdom}

\author{G J Rodgers\footnote{Electronic address: G.J.Rodgers@brunel.ac.uk}}

\address{Department of Mathematics and Statistics,Brunel University,
Uxbridge, Middlesex, UB8~3PH, United Kingdom}

\date{March 9, 1998}

\maketitle

\begin{abstract}The Bak--Sneppen model is shown to fall
into a different universality class with the introduction
of a preferred direction, mirroring the situation in spin
systems. This is first demonstrated by numerical simulations
and subsequently confirmed by analysis of the multi--trait
version of the model, which admits exact solutions in the
extremes of zero and maximal anisotropy.
For intermediate anisotropies, we show that the spatio\-temporal
evolution of the avalanche has a power law ``tail'' which passes
through the system for any non--zero anisotropy but remains fixed
for the isotropic case, thus explaining the cross\-over in behaviour.
Finally, we identify the maximally anisotropic model
which is more tractable and yet more generally applicable
than the isotropic system.
\end{abstract}

\pacs{64.60.Lx, 05.40.+j, 05.70.Jk}

Accepted for publication in J. Phys. A

\maketitle
\begin{multicols}{2}
\narrowtext

%%
%%                  INTRODUCTION
%%
%%  Background (original model, M=infinity solution)
%%  Justification (gravity in granular compaction, asymmetry
%%  in foodwebs). Only study abstract model.
%%
\section{Introduction}
\label{sec:intro}

The Bak--Sneppen model was originally introduced as a crude caricature
of biological macro\-evolution in an attempt to explain the
distribution of extinction sizes observed in the
fossil record~\cite{BS,BSRev,newman}.
Although still widely studied in this context,
there is also a great deal of interest in analysing
the model from a purely abstract viewpoint.
This is because it is currently the
simplest and most tractable of the class of
{\em extremal dynamical} models,
which themselves form a subset of
self--organised critical systems~\cite{BSRev,sol2,sol1}.
Extremal dynamical models are so called because they are driven by
the the selection of some globally extremal value which
dynamically interacts with nearby sites.
They naturally evolve towards a ``critical state''
(a second order phase transition)
without any characteristic length or time scales.

It might appear that the Bak--Sneppen model is well suited
to adopt the r\^{o}le of the ``Ising model''
of extremal dynamical systems.
We believe that this is not the case, and in this paper
we detail an even simpler version of the model which
is more open to analysis whilst retaining all
the essential behaviour of the original.
The inspiration behind this new model can be most
clearly described by analogy with spin systems~\cite{yeomans}.
The Heisenberg spin model is isotropic because
the spin vectors have no preferred direction.
However, when even the slightest anisotropy is introduced,
a preferred direction is created and the system
falls into a different universality class.
Furthermore, this is the same class
as the highly anisotropic Ising model,
where the spin vectors can only lie parallel to the direction
of quantisation.
So not only is the Ising model in some sense more general
than the Heisenberg model, it is also simpler and hence
more tractable.

The original incarnation of the Bak--Sneppen model
is like the Heisenberg model in that it too is isotropic.
If the analogy with spin systems is to hold true, then
the introduction of anisotropy into the Bak--Sneppen model
should result in a different universality class.
We find that this is indeed the case, at least for one
dimensional systems, and conclude that, unless there
is some reason for assuming perfect isotropy,
it is the anisotropic model that should be treated as the general case
and the isotropic version as a special limiting instance.
It is also possible to identify a {\em maximally anisotropic}
Bak--Sneppen model which may serve as
the true analogue of the Ising model for extremal dynamical systems.
We postpone until Sec.\ref{sec:disc} the question of whether
isotropy should be assumed in any known application
of the model.

This paper is organised as follows.
Numerical simulations of anisotropic systems are described in
Sec.~\ref{sec:model} and the exponents for the new universality
class are given.
By switching to the multi--trait model, a full solution
of the maximally anisotropic system is found which explicitly
demonstrates the cross\-over to the new class.
This solution is derived in Sec.~\ref{sec:multi} alongside
the known result for the isotropic model.
An exact solution for intermediate anisotropies was not
forthcoming, but by employing an alternative means of analysis
it is possible to show that this new class also applies
to any non--zero anisotropy.
This is presented in Sec.~\ref{sec:arb_anis}.
Finally, in Sec.~\ref{sec:disc} we discuss the applicability
of this new class in real situations,
and consider the potential of the maximally anisotropic
model in future analytical treatments.

%%
%%              THE ANISOTROPIC MODEL
%%
%%  Definition of the standard model and a brief review of
%%  its behaviour, then the anisotropic variant and its
%%  results. M=1 throughout.
%%
\section{The Anisotropic Model}
\label{sec:model}

Before coming to consider anisotropy
we briefly summarise the isotropic model and some
of its known results~\cite{BS,BSRev}.
$N$~scalars $f_{i}$\,,
where \mbox{$i=1\ldots N$}\,,
are placed on a one--dimensional
lattice with periodic boundary conditions.
The~$f_{i}$\,, known as ``barriers,''
are random numbers uniformly distributed
on $[0,1]$, although the system behaves in essentially
the same manner regardless of the particular choice of distribution.
At each time step the global minimum of all the $f_{i}$
is found, and it and its two nearest neighbours are
given new random values from the same distribution as before.
This process is then repeated {\em ad infinitum}.

Despite such minimalist dynamics the model exhibits
a rich variety of non--trivial behaviour.
It evolves towards a statistical steady state
in which the bulk of the $f_{i}$ are uniformly
distributed on $[f_{c},1]$, where
the threshold value $f_{c}$ is a function of the
lattice dimension and connectivity.
For the one--dimensional lattice considered here,
\mbox{$f_{c}\approx0.667$}.
A finite number of barriers form a tail on $[0,f_{c}]$
and it is in this tail that the global minimum is
always to be found.
Both the spatial and temporal correlation
functions are power law in form, signifying the
existence of a critical state with no characteristic
length or time scales.
The distribution for
the absolute distance between successive minima $\Delta x$
is

\begin{equation}
P_{\rm JUMP}(\Delta x)\sim(\Delta x)^{-\pi}\:,
\end{equation}

\noindent{}where $\pi=3.23\pm0.02$.
The probability that the minimum is at the same site
at times $t_{0}$ and $t+t_{0}$ is given by

\begin{equation}
P_{\rm ALL}(t) \sim t^{-\tau_{\rm ALL}}\:,
\end{equation}

\noindent{}with $\tau_{\rm ALL}=0.42\pm0.02$.
This holds true as long as \mbox{$t\ll t_{0}$} and
ageing effects can be ignored~\cite{Age1,Age2}.

The model defined above is isotropic because the
interaction between the global minimum and the other
barriers is the same in both directions.
In other words, if the current minimum is~$f_{i}$ then
barriers~$f_{i-1}$, $f_{i}$
and $f_{i+1}$ are reset,
so the minimum is just as likely to
jump to the left as it is to the right.
Consider what happens when the rules are altered
so that $f_{i-1}$, $f_{i}$ and $f_{i+2}$ are reset instead.
The system now has an inherent bias to the right
and we would expect an avalanche to be more
likely to propagate in that direction.
This constitutes an {\em anisotropic} model
since there now exists a preferred direction
for the global minimum to drift.

\begin{figure}
\centerline{\psfig{file=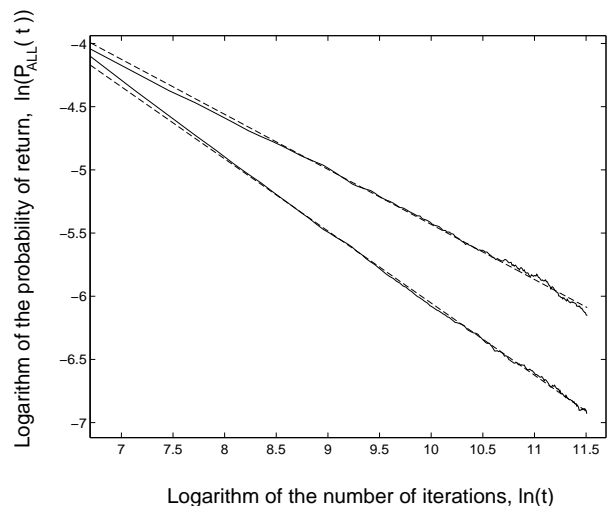,width=3.5in}}
\caption{A log--log plot of $P_{\rm ALL}(t)$, the
probability for the active
site to return to its original position after a time~$t$.
The upper line is from the standard isotropic model and
the lower line is from an anisotropic system in which the barriers
$f_{i-1}$, $f_{i}$ and $f_{i+3}$ are reset at every time step,
where $f_{i}$ is the current global minimum.
The dashed lines have slopes of $-0.42$
and $-0.59$, respectively.
The data for the anisotropic system has been moved upwards to
allow for direct comparison with the isotropic case.
The simulations were performed on an \mbox{$N=10^{4}$}
lattice, for \mbox{$5\times10^{3}N$} iterations in the isotropic
case and \mbox{$5\times10^{4}N$} in the anisotropic case.
}
\label{f:p_all}
\end{figure}

We have performed extensive numerical simulations of the
anisotropic model and have observed that the system behaves
in qualitatively the same manner as the isotropic model.
However,
the correlation distributions $P_{\rm JUMP}$ and $P_{\rm ALL}$
have different exponents,
\mbox{$\pi^{\uparrow}=2.42\pm0.05$} and
\mbox{$\tau^{\uparrow}_{\rm ALL}=0.59\pm0.03$},
so the system falls into a different universality class
to the isotropic case.
Plots of $P_{\rm ALL}$ for both classes are given
in Fig.~\ref{f:p_all} for direct comparison.
$P_{\rm JUMP}$~is uniformly lower for jumps against the
direction of anisotropy as for jumps with it, but
the same exponent applies in both directions.
The threshold value $f_{c}$ also drops,
but this is simply due to the increased spreading out
of the avalanche and has nothing to do with
the loss of isotropy.

The new universality class is not just restricted to
this one example.
Simulations have shown that if barriers
$f_{i-a}$, $f_{i}$ and $f_{i+b}$ are reset,
where $a$ and~$b$ are arbitrary positive integers,
then the same class holds for {\em any} \mbox{$a\neq b$}.
The dynamics can be further generalised by considering
ranges of sites on each side of the minimum, and either selecting
all of these sites or just a random sample.
Here, anisotropy corresponds to a larger range
on one side than on the other.
In all cases the same exponents are found for any non--zero anisotropy,
although convergence can be very slow when
the anisotropy is weak,
a point that will be explicitly demonstrated for the multi--trait model
in Sec.~\ref{sec:arb_anis}.

\vspace{0.7in}

%%
%%             THE MULTI-TRAIT MODEL
%%
%%  Definition of the multi-trait model and its solution
%%  for M=infinity. Solution of the (0,1) case for the same.
%%
\section{The Multi--Trait Model}
\label{sec:multi}

The consequences of introducing anisotropy
into the Bak--Sneppen model
can be more fully investigated by switching
to the multi--trait framework~\cite{multi}.
In the multi--trait model each site has $M$
internal degrees of freedom, that is $M$~different
barriers rather than just the usual~1.
At each time step the smallest of all the \mbox{$N\times M$}
barriers in the system is found and reset.
One of the $M$ barriers from each of its neighbouring sites is
selected at random and also reset,
so three barriers are reset in total.
Then the new global minimum is found and the
process is iterated indefinitely.
For finite $M$ the system belongs to the same universality class
as the standard $M=1$ model, but for \mbox{$M\rightarrow\infty$}
it falls into a different class and, furthermore, can be solved
exactly.
To see why this is so, we must first define
what is meant by a
\mbox{$\lambda$--{\em avalanche}}.

For any given value of \mbox{$\lambda<f_{c}$}
the global minimum can be either greater or less
than~$\lambda$.
Hence a time series of the minimum will consist
of regions where it is less than~$\lambda$ alternating
with regions where it is greater than~$\lambda$.
Each block for which the global minimum is less that $\lambda$
is defined as a \mbox{$\lambda$--avalanche}.
During a \mbox{$\lambda$--avalanche}
any barrier smaller than $\lambda$ is called
{\em active} since the avalanche cannot finish until all
of the active barriers have been made inactive, that is
when they have all been reset to values greater than~$\lambda$.
There are only two ways in which a barrier
can get reset,
it either becomes the global minimum
or belongs to an adjacent site to
the minimum and is selected with probability~\mbox{$1/M$}.
However, the latter possibility cannot occur in
the \mbox{$M\rightarrow\infty$} limit since there
are only a finite number of active barriers
in the system at any one time, so
the probability of selecting one at random is vanishingly small.
Hence each active barrier must eventually become
the global minimum and it will then initiate
a sub--avalanche that can change inactive barriers to active,
but never the other way around.
Furthermore, since the sub--avalanches from different active barriers
propagate independently of each other,
the active barriers can be reset in any order
and there is no longer any need to keep track of
which is actually the global minimum.

The temporal correlations for \mbox{$M\rightarrow\infty$}
are the same as for a mean field model
in which the neighbours of
the minimum are chosen at random,
so the introduction of anisotropy will make no difference.
Rather than repeat the analysis here, we simply quote the main
result and refer the reader to~\cite{multi} for
details of the derivation.
If $P_{\lambda}(t)$ is the probability that a $\lambda$--avalanche
lasts for exactly time~$t$, then

\begin{equation}
P_{\lambda}(t)\sim
t^{-3/2}\;G(\;t\,(\lambda-\mbox{$\frac{1}{2}$})^{2}\,)
\label{e:time}
\end{equation}

\noindent{}as \mbox{$\lambda\rightarrow f_{c}=\frac{1}{2}$},
where $G(x)$ is a scaling function that
tends to a constant value for \mbox{$x\rightarrow0$}.
As expected, $P_{\lambda}$ has the usual mean field exponent
of~$\frac{3}{2}$.
Since~(\ref{e:time}) holds independently of the
spatial structure of the system, we can already conclude that
the threshold barrier value~$f_{c}$ will be~$\frac{1}{2}$
regardless of the degree of anisotropy.

Anisotropy will clearly effect the spatial correlations and
so we present the following analysis in some detail, starting
with the isotropic model.
As in the derivation of~(\ref{e:time}) the algebra
is simplified by stipulating that the active barrier
always takes the value of~1 when it is reset.
This makes no qualitative difference to the results.
Hence the central barrier is always
made inactive, but the barriers reset in each of the adjacent
sites may become active with probability~$\lambda$.
Let $g_{r}$ denote the probability that resetting an active
barrier at the origin causes at least one of the barriers
in site~$r$ to become active.
Then \mbox{$1-g_{r}$} is the probability that no barriers
become active, which can be related to \mbox{$1-g_{r-1}$}
and \mbox{$1-g_{r+1}$} by the difference equation

\begin{eqnarray}
1-g_{r}&=&(1-\lambda)^{2}+\lambda^{2}(1-g_{r-1})(1-g_{r+1})\nonumber\\
&&\mbox{}+\lambda(1-\lambda)\left\{(1-g_{r-1})+(1-g_{r+1})\right\} \:.
\label{e:n_diff}
\end{eqnarray}

\noindent{}This can be derived by considering what happens
when an active barrier at the origin is reset.
The probability of creating no new active barriers
is \mbox{$(1-\lambda)^{2}$}, in which case the avalanche
will end and site~$r$ will definitely not become active.
This is catered for by the first term on the right hand side
of~(\ref{e:n_diff}).
Similarly, the second and third terms account for the
creation of active barriers in one or both of the adjacent sites,
which may subsequently propagate to site~$r$
with probabilities $g_{r-1}$ and $g_{r+1}$, assuming
$g_{r}$ to be translationally invariant.
(\ref{e:n_diff})~can be rearranged to give

\begin{equation}
g_{r} = \lambda(g_{r-1}+g_{r+1})-\lambda^{2}g_{r-1}g_{r+1}\:.
\label{e:g_r}
\end{equation}

\noindent{}If the whole \mbox{$\lambda$--avalanche} starts from
a single active barrier at~\mbox{$r=0$}, then \mbox{$g_{0}=1$}
and (\ref{e:g_r}) can be solved to give

\begin{equation}
g_{r} = \frac{12}{(r+3)(r+4)}
\label{e:exactis}
\end{equation}

\noindent{}for $\lambda=\frac{1}{2}$\,,
explicitly demonstrating the asymptotic power law
behaviour~\mbox{$g_{r}\sim1/r^{2}$}.

\begin{figure}
\centerline{\psfig{file=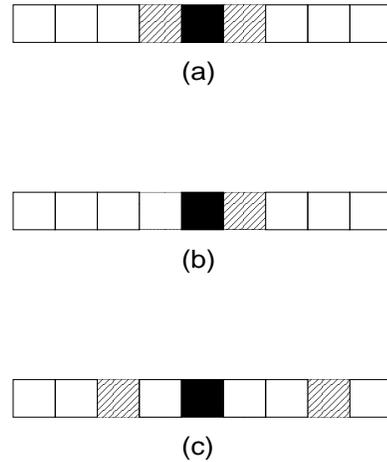,width=2in}}
\caption{Examples of various definitions of anisotropy.
In all cases the solid square is
the site with the global minimum and the shaded squares are the other
sites in which a barrier is also reset.
The shaded sites are $a$ places to the left of the
site with the minimum, and $b$ places to its right.
(a)~The standard model~\mbox{$a=b=1$}.
(b)~The maximally anisotropic model~\mbox{$a=0$}, \mbox{$b=1$}.
(c)~An intermediate case~\mbox{$a=2$}, \mbox{$b=3$}.}
\label{f:ab_models}
\end{figure}

There are many ways in which anisotropy could be incorporated
into this framework, but for clarity we restrict our
attention to just a single definition.
At every time step the global minimum barrier is found,
say in site~$i$, and reset.
The anisotropic interaction consists of randomly selecting one of
the $M$ barriers in each of the sites~$i-a$ and $i+b$
and resetting them both,
where the parameters $a$ and $b$ are positive integers.
Some examples are given in Fig.~\ref{f:ab_models}.
Note that if $a$ and $b$ share a common factor, say $c$, then the
system will trivially decouple into $c$ independent sublattices.
For instance, if \mbox{$a=b=2$} then all
the even numbered sites will decouple from all the odd
numbered sites and the two sublattices will evolve independently
of each other.
Thus we can safely assume that $a$ and $b$ are coprime.
As a corollary any system with \mbox{$a=b$} is equivalent
to the standard model \mbox{$a=b=1$}.
Similarly, if $a$ is equal to zero we can
take \mbox{$b=1$} without loss
of generality, and vice versa
if~\mbox{$b=0$}.

The maximally anisotropic system
with \mbox{$a=0$} and \mbox{$b=1$} can be solved
in much the same way as the isotropic case.
Since only two barriers are reset at every
time step anyway there is no need to set
the central barrier to~1 as before.
The resulting difference equation is similar to (\ref{e:n_diff})
and can be derived in an entirely analogous manner,

\begin{eqnarray}
1-g_{r}&=&(1-\lambda)^{2}+\lambda^{2}(1-g_{r})(1-g_{r+1})\nonumber\\
&&\mbox{}+\lambda(1-\lambda)\left\{(1-g_{r})+(1-g_{r+1})\right\} \:,
\end{eqnarray}

\noindent{}which rearranges to

\begin{equation}
g_{r}=\frac{\lambda g_{r-1}}{1-\lambda+\lambda^{2}g_{r-1}}\:.
\label{e:g_01}
\end{equation}

\noindent{}For \mbox{$\lambda=\frac{1}{2}$} this admits
the exact solution

\begin{equation}
g_{r} = \left\{
	  \begin{array}{ll}
	    \frac{2}{2+r} & \mbox{ for $r\geq0$, and} \\
	    0 & \mbox{for $r<0$,}
          \end{array}
	\right.
\label{e:maxanis}
\end{equation}

\noindent{}so now $g_{r}\sim1/r$ for large~$r$,
giving a power law with an exponent of~$1$.

An exact expression for 
\mbox{$\lambda\neq\frac{1}{2}$} can also be
found by substituting \mbox{$g_{r}=1/z_{r}$}
into~(\ref{e:g_01}). This gives a linear difference
equation for the~$z_{r}$,

\begin{equation}
z_{r}=\frac{1-\lambda}{\lambda}\;z_{r-1}+\lambda\:,
\end{equation}

\noindent{}which can be solved to give

\begin{equation}
z_{r}=\frac{\lambda^{2}}{1-2\lambda}
\left\{\left(\frac{1-\lambda}{\lambda}\right)^{r+2}-1\right\}\:.
\end{equation}

\noindent{}For $\lambda<\frac{1}{2}$, $z_{r}$ blows up exponentially
in $r$ and so $g_{r}$ will exponentially {\em decay} to zero.
If \mbox{$\lambda>\frac{1}{2}$} then $g_{r}$ will exponentially
decay to a constant value for large~$r$, corresponding to a
non--zero probability of initiating an infinite avalanche.
However, this latter case is of academic interest only since the
underlying simplification of the \mbox{$M\rightarrow\infty$}
limit rests on there being only a finite number of
active barriers at any one time,
which is no longer true when \mbox{$\lambda>\frac{1}{2}$}\,.

The exponent for $g_{r}$ is related to
the exponent for~$P_{\rm JUMP}$ by~\mbox{$\pi=\tau_{\rm R}+1$},
where \mbox{$g_{r}\sim r^{-\tau_{\rm R}}$} and
\mbox{$P_{\rm JUMP}(\Delta x)\sim(\Delta x)^{-\pi}$}.
Hence the analysis given above demonstrates that $\pi$
changes from~3 to~2 with the introduction of anisotropy.
This should be compared to the numerical results in
Sec.~\ref{sec:model} for \mbox{$M=1$} systems, where $\pi$
went from $3.23\pm0.02$ to~$2.42\pm0.05$.
In both cases the exponent jumps in the same direction and
by a roughly similar amount.
Furthermore, for~\mbox{$M\rightarrow\infty$}
the exponent for~$P_{\rm ALL}(t)\sim t^{-\tau_{\rm ALL}}$
obeys~\mbox{$\tau_{\rm ALL}=(2\tau_{\rm R})^{-1}$}.
Hence $\tau_{\rm ALL}$
increases from $\frac{1}{4}$ to~$\frac{1}{2}$,
and again a similar jump was observed for~\mbox{$M=1$},
where $\tau_{\rm ALL}$ increased from
$0.42\pm0.02$ to $0.59\pm0.03$.
Thus the change in behaviour in the \mbox{$M\rightarrow\infty$} limit
is also representative of the \mbox{$M=1$} case.

\vspace{0.5in}

%%
%%             ARBITRARY ANISOTROPY
%%
%%  Description of the independent avalanching lattice and
%%  relation to the actual lattice. Solutions (linear for
%%  the broadening, continuum approx for the power-law decay)
%%  and explanation of new exponents.
%%
\section{Arbitrary Anisotropy}
\label{sec:arb_anis}

It remains to be seen whether systems with intermediate
anisotropies do indeed fall into the same universality class as the
maximally anisotropic model, as implied by the analogy
with the Heisenberg and Ising spin models.
Unfortunately, the style of analysis adopted in the previous section is of
little use here since the difference equation (\ref{e:g_r})
admits no straightforward solutions for arbitrary $a$ and~$b$.
The difficulty stems from the fact that
the interactions are now between non--adjacent sites.
One way around this problem is to
find a separate lattice representation for the avalanche
in which only nearest neighbours interact.
This could then be mapped onto the one--dimensional
substrate in such a way that nearest neighbours
on the avalanche lattice map onto interacting sites
on the substrate.

To do this unambiguously, it is necessary to employ
a two--dimensional lattice $(n,m)$ which
represents the entire spatio\-temporal extent of the avalanche.
The mapping from sites $(n,m)$ on the avalanche lattice
to sites $r$ on the one--dimensional substrate
is derived as follows.
The origin $(0,0)$ corresponds to \mbox{$r=0$}.
Any given site $(n,m)$ can be reached
by taking $n$ steps to the left and
$m$ steps to the right, in any order.
For arbitrary $a$ and~$b$, the resulting value of $r$ is

\begin{equation}
r=na-mb\:.
\label{e:r_namb}
\end{equation}

\noindent{}Each value of $r$ corresponds to the set of
points $(n_{i},m_{i})$ that obey~(\ref{e:r_namb}).
Successive points are separated by the constant
displacement vector

\begin{eqnarray}
\Delta_{nm}&=&(n_{i+1},m_{i+1})-(n_{i},m_{i}) \nonumber\\
&=&(n_{i+1}-n_{i},m_{i+1}-m_{i}) \nonumber\\
&=&(b,a)\:. 
\label{e:delta_nm}
\end{eqnarray}

\begin{figure}
\centerline{\psfig{file=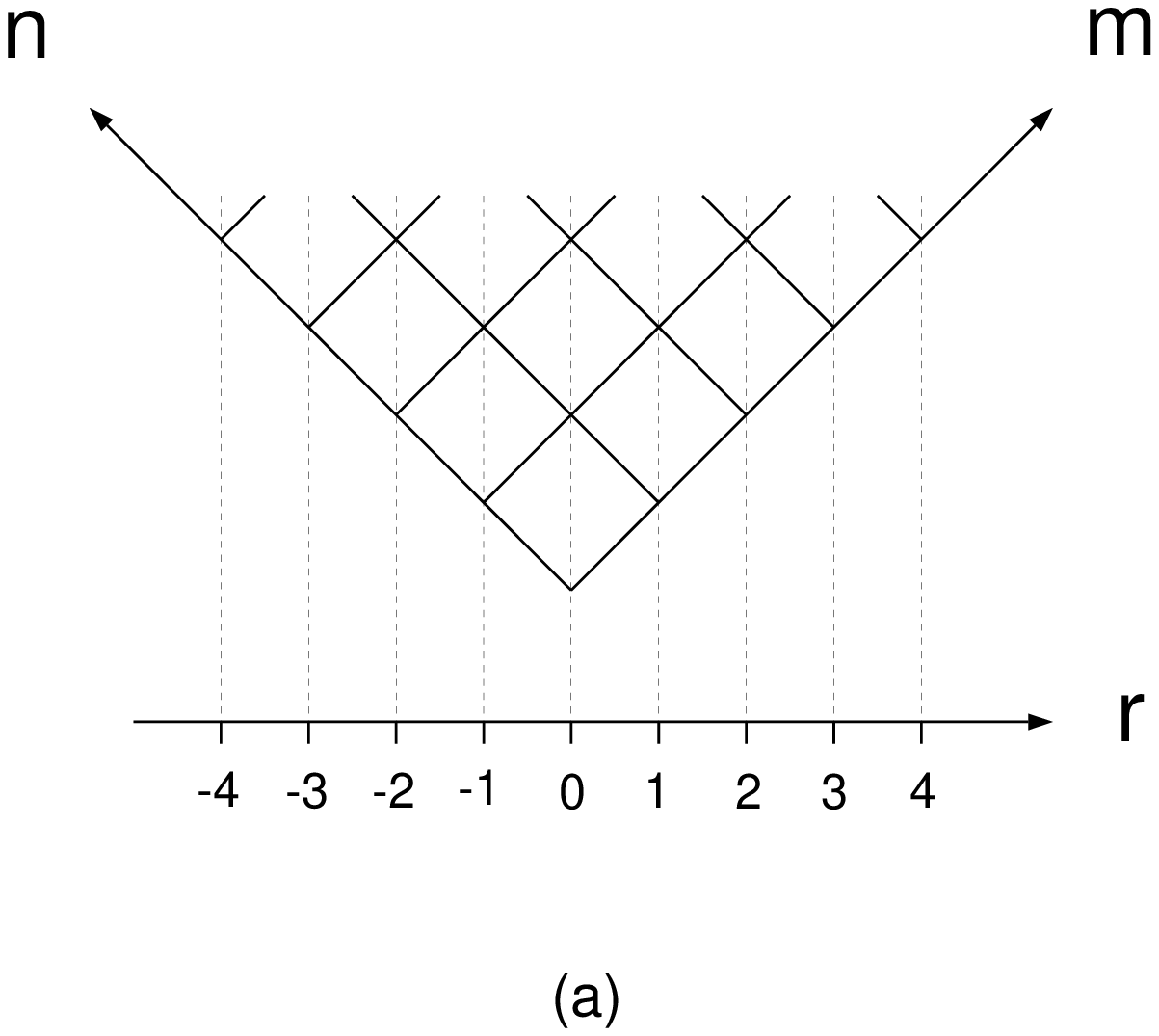,width=1.8in}}
\centerline{\psfig{file=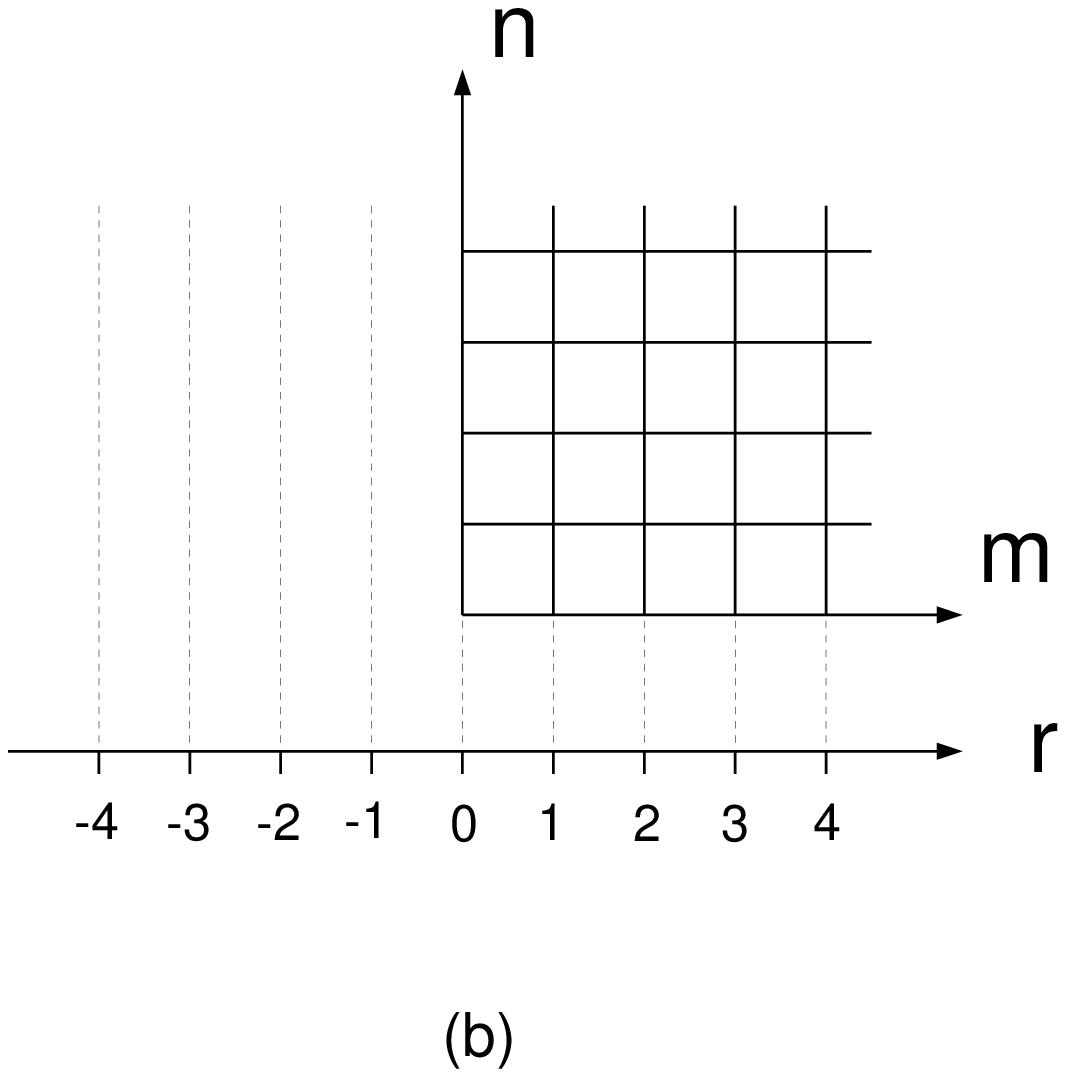,width=1.8in}}
\centerline{\psfig{file=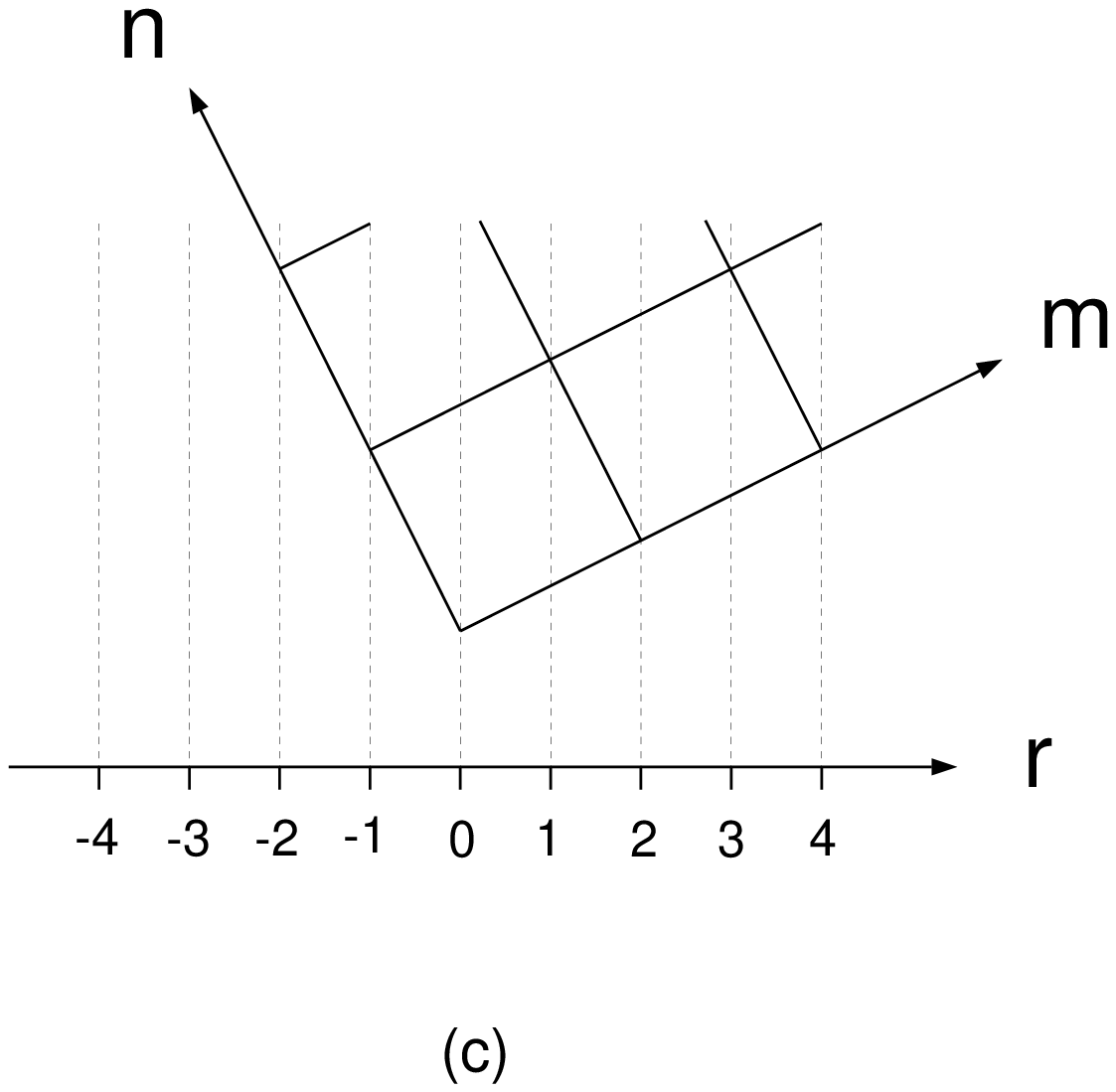,width=1.8in}}
\caption{The projection from the
$(n,m)$ lattice to sites $r$ on
the one-dimensional substrate.
Adjacent sites on the $(n,m)$ lattice correspond
to sites~$r$ that interact.
(a)~The isotropic
model \mbox{$a=b=1$}. The dotted lines connect
all the $(n,m)$ that correspond to the same value of~$r$.
(b)~The maximally anisotropic case \mbox{$a=0$}, \mbox{$b=1$}.
(c)~An intermediate case \mbox{$a=2$}, \mbox{$b=1$}.}
\label{f:nm_lattice}
\end{figure}

\noindent{}That $\Delta_{nm}$ is the smallest displacement vector
follows from the coprime nature of $a$ and~$b$.
The mapping from $(n,m)$ to $r$ can thus be regarded
as a {\em projection} from the two--dimensional
avalanche lattice to the one--dimensional substrate.
An example is given in Fig.~\ref{f:nm_lattice}(a)
for the isotropic model.
When \mbox{$a\neq b$} the \mbox{$(n,m)$--lattice}
becomes rotated relative to the projection lines,
so for instance when \mbox{$a=0$} and \mbox{$b=1$} the
lattice lies completely on its side, as in Fig.~\ref{f:nm_lattice}(b).
An intermediate case is
given in Fig.~\ref{f:nm_lattice}(c).

To quantify this relationship further,
let $p_{nm}$ be the probability that site $(n,m)$ is active.
Site $r$ will remain inactive throughout the entire avalanche
only if all of its corresponding $(n,m)$ are also inactive, so

\begin{equation}
g_{r}=1-\prod_{n,m}(1-p_{nm})\:,
\label{e:prod}
\end{equation}

\noindent{}where the product is taken over all the
$n$ and $m$ that obey~(\ref{e:r_namb}).
The advantage of this approach is that
varying the anisotropy only effects {\em which} $p_{nm}$
contribute to~(\ref{e:prod}), the $p_{nm}$
themselves are {\em entirely unaltered}.
Thus a unique solution to the $p_{nm}$ exists
which, if found, could be applied to any anisotropy
through~(\ref{e:prod}) without modification.

The next step is to find the solution for the~$p_{nm}$.
Each site $(n,m)$ has $M$ barriers which, if active,
may create active barriers in
either or both of sites $(n+1,m)$ and $(n,m+1)$.
Since we are still in the \mbox{$M\rightarrow\infty$} limit,
the sub--avalanches initiated by different active barriers are
independent and can
be arranged so as to form a compact avalanche
on the $(n,m)$ lattice.
A site $(n,m)$ can only become active if one of its
barriers is reset to a value less that~$\lambda$
due to the interaction with an active neighbouring site.
The neighbours in question are the two diagonally lower sites
\mbox{$(n-1,m)$} and \mbox{$(n,m-1)$},
so the probability of either event occurring independently
is \mbox{$\lambda\,p_{n-1m}$}
and~\mbox{$\lambda\,p_{nm-1}$}, respectively.
Thus the difference equation for the $p_{nm}$ is

\begin{equation}
p_{nm}=\lambda(p_{n-1m}+p_{nm-1})-\lambda^{2}p_{n-1m}p_{nm-1}\:.
\label{e:diff_pnm}
\end{equation}

By dropping the second term on the right hand side
of~(\ref{e:diff_pnm}),
a linear equation is obtained which has the exact solution

\begin{equation}
p^{\rm lin}_{nm}=\frac{(n+m)!}{n!\,m!}\,\lambda^{n+m}\:.
\label{e:pnm_lin}
\end{equation}

\noindent{}Except for a missing normalisation
factor of \mbox{$2^{-(n+m)}$},
the coefficients \mbox{$(n+m)!/(n!\,m!)$}
describe a binomial distribution with equal
probability of either outcome.
For large $n$ and $m$ this binomial is well approximated
by a Gaussian distribution with mean \mbox{$(n+m)/2$} and
variance \mbox{$(n+m)/4$}\,,

\begin{equation}
p^{\rm lin}_{nm}\approx\sqrt{\frac{2}{\pi(n+m)}}\,
\exp\left\{{-\frac{1}{2}\frac{(n-m)^{2}}{n+m}}\right\}\,
(2\lambda)^{n+m} \:.
\label{e:lin_sol}
\end{equation}

\noindent{}This expression is vanishingly small
except for points that lie near the line~\mbox{$n=m$},
which form a non--vanishing ``tail''.
A cut through points of equal \mbox{$n+m$}
shows that this tail has a Gaussian cross section
of width \mbox{$\frac{1}{2}\sqrt{n+m}$}\,,
so it becomes broader down its length.
The behaviour down the
centre of the tail depends upon the value of~$\lambda$.
For \mbox{$\lambda\neq\frac{1}{2}$},
$p^{\rm lin}_{nn}$ either blows up or decays
exponentially according to the factor of $(2\lambda)^{2n}$
in~(\ref{e:lin_sol}).
At the critical point \mbox{$\lambda=\frac{1}{2}$} this factor
becomes unity and instead the tail
exhibits power law decay
\mbox{$p^{\rm lin}_{nn}\sim n^{-1/2}$}.

Numerical integration of the full difference equation~(\ref{e:diff_pnm})
shows that the exact solution of $p_{nm}$
does indeed have a Gaussian tail of variance \mbox{$(n+m)/4$} which
decays as a power law for \mbox{$\lambda=\frac{1}{2}$},
in agreement with the expression for $p^{\rm lin}_{nm}$.
However, the exponent for the power law decay is different in both cases,
\mbox{$p_{nn}\sim n^{-1}$} for the exact solution as
opposed to \mbox{$p^{\rm lin}_{nn}\sim n^{-1/2}$}.
The correct exponent can be recovered by restoring the non-linear
term in (\ref{e:diff_pnm}) and instead considering the
equivalent continuum approximation~\cite{ctm}.
Let $x$ and $y$ be continuous variables corresponding to
$n$ and~$m$, and define \mbox{$h(x,y)=p_{nm}$}.
Using this notation,

\begin{eqnarray}
\nabla h(x,y)
&=&\frac{\partial h(x,y)}{\partial x}+\frac{\partial h(x,y)}{\partial y}
\nonumber\\
&\approx &(p_{nm}-p_{n-1m})+(p_{nm}-p_{nm-1}) \:,
\end{eqnarray}

\noindent{} and (\ref{e:diff_pnm}) can be rewritten as

\begin{equation}
\frac{1}{2}\nabla h=(2\lambda-1)h-\lambda^{2}h^{2}\:.
\end{equation}

\noindent{}This can be simplified by making
the substitution \mbox{$h(x,y)=1/z(x,y)$}
and the change of variables \mbox{$u=x+y$} and \mbox{$v=x-y$},
giving

\begin{equation}
\frac{\partial z}{\partial u} = (1-2\lambda)\,z+\lambda^{2} \:.
\label{e:a}
\end{equation}

For \mbox{$\lambda=\frac{1}{2}$}
the first term on the right hand side
of (\ref{e:a}) vanishes and straightforward integration gives

\begin{equation}
z(u,v)=\frac{u}{4}+A(v)\:,
\end{equation}

\noindent{}where $A(v)$ is an arbitrary function of~$v$
that is found from the boundary conditions.
There are two sets of boundary conditions, one for the line
\mbox{$m=0$} and another for the line~\mbox{$n=0$}.
It is clear from (\ref{e:diff_pnm}) that
\mbox{$p_{n0}=\lambda^{n}$} exactly.
The line \mbox{$m=0$} is the same as the line \mbox{$y=0$},
which maps onto \mbox{$u=v$} after the change of variables,
so the first boundary condition is \mbox{$z(u,u)=\lambda^{-u}$}.
Similarly, \mbox{$p_{0m}=\lambda^{n}$} and
the line \mbox{$n=0$} corresponds to \mbox{$u=-v$},
so the second boundary condition is
\mbox{$z(u,-u)=\lambda^{-u}$}.
This allows for $A(v)$ to be fixed and the full solution is

\begin{equation}
z(u,v)=\frac{u-|v|}{4}+2^{|v|}\:.
\end{equation}

\noindent{}Along the tail \mbox{$v=0$},
\mbox{$z(u,0)\sim u$} and so
\mbox{$h(x,x)\sim x^{-1}$}, giving the correct
exponent for the decay.
However, moving away from the tail results
in exponential growth in $z(u,v)$ and hence exponential
decay in~$h(x,y)$.
Thus the continuum approximation predicts the correct
exponent for the decay of the tail but the wrong
cross sectional shape, that is, exponential rather than Gaussian.

The solution for \mbox{$\lambda<\frac{1}{2}$} can be found
by following exactly the same procedure.
This results in

\begin{equation}
z(u,v)=\frac
{\left( (1-2\lambda)\lambda^{-|v|}+\lambda^{2}\right)
e^{(1-2\lambda)(u-|v|)}-\lambda^{2}}
{1-2\lambda}\:.
\label{e:a_full}
\end{equation}

\noindent{}As before, this expression increases exponentially
away from the tail. For~\mbox{$v=0$} it reduces to

\begin{equation}
z(u,0)=\frac{
(1-\lambda)^{2}e^{(1-2\lambda)u}-\lambda^{2}}
{1-2\lambda} \:,
\end{equation}

\noindent{}which blows up exponentially in~$u$.
Hence $h(x,x)$ decays to zero when \mbox{$\lambda<\frac{1}{2}$}\,,
giving an exponential cut-off in
the distribution of avalanche sizes.
We note in passing that (\ref{e:a_full}) also holds
for \mbox{$\lambda>\frac{1}{2}$} but, as explained in the
Sec.~\ref{sec:multi}, such values of $\lambda$ bear no
relevance to actual systems.

Armed with the solution to the $p_{nm}$\,, we can now
derive the exponents of~$g_{r}$
for when \mbox{$\lambda=\frac{1}{2}$}.
First consider the isotropic case \mbox{$a=b=1$}.
Under the projection in Fig.~\ref{f:nm_lattice}(a)
all the points down the centre of the tail
are mapped onto the origin, so
$g_{r\neq0}$ will only start to receive
non-vanishing contributions once
the tail has become sufficiently wide.
Since the tail broadens like~\mbox{$(n+m)^{1/2}$}
the first $p_{nm}$ to contribute to $g_{r}$
will lie on the line \mbox{$n+m\sim r^{2}$},
by which point the tail will have already decayed to
\mbox{$p_{nm}\sim(n+m)^{-1}\sim r^{-2}$}.
Once these $p_{nm}$ are substituted into
the infinite product in~(\ref{e:prod}),
the leading order terms in $g_{r}$ will look something like

\begin{equation}
g_{r}\approx\frac{a_{2}}{r^{2}}+\frac{a_{4}}{r^{4}}+\frac{a_{6}}{r^{6}}+
\ldots\:\:\:,
\label{e:g_r2}
\end{equation}

\noindent{}where the $a_{i}$ are constants.
Hence \mbox{$g_{r}=O(1/r^{2})$}, in agreement
with the exact solution~(\ref{e:exactis}).

The situation is very different in an anisotropic
system~\mbox{$a\neq b$}.
As can be seen in Figs.~\ref{f:nm_lattice}(b)
and~\ref{f:nm_lattice}(c), the tail is no longer
vertical but cuts through the projection lines
at a finite angle, passing over all \mbox{$r\geq0$}
(the preferred direction is to the right in both of these examples).
Furthermore, since the gap between successive
$p_{nm}$ mapped onto the same $r$ is finite,
and the tail broadens without limit, then for sufficiently large
$r$ an arbitrarily large number of $p_{nm}$ will contribute
to each~\mbox{$g_{r}$}.
Each of these $p_{nm}$ will be proportional to~\mbox{$1/r$}, so
the analogous expression to~(\ref{e:g_r2}) will be
\mbox{$g_{r}=O(1/r)$}
and power law behaviour with an exponent of~$1$ is recovered,
in agreement with (\ref{e:maxanis}) and numerical simulations.
For~\mbox{$r<0$} only exponentially small $p_{nm}$
contribute and $g_{r<0}$
takes some exponentially decaying form.

Thus the cross\-over in behaviour from the anisotropic to
the isotropic model in the \mbox{$M\rightarrow\infty$} limit
can be attributed to the difference between
a power law tail that moves across the substrate, and one whose
centre is fixed and can only broaden at a much slower rate.
The convergence to the new behaviour can be very slow, especially
for weak anisotropy~\mbox{$a\approx b$}.
Indeed, since \mbox{$g_{r}\sim1/r+O(1/r^{2})$} the rate of convergence
is itself a power law.
Slow convergence was also observed in the simulations
of the \mbox{$M=1$} models is Sec.~\ref{sec:model}.
However, for \mbox{$M=1$} the spatial correlations
were power law in both directions, whereas in the
\mbox{$M\rightarrow\infty$} limit the correlations
{\em against} the direction of anisotropy decay exponentially.
This difference presumably arises because the sub--avalanches initiated
from different active sites are
no longer independent for finite~$M$.

%%
%%                 DISCUSSION
%%
%%  Possible future directions, ie. solution of (0,1) model
%%  with M=1, relationship between the two exponents.
%%
\section{Discussion}
\label{sec:disc}

It should come as no surprise that the anisotropic
Bak--Sneppen model has different critical
exponents to its isotropic equivalent.
Universality classes depend upon
the dimensionality and symmetries of the model in question,
so the loss of symmetrical interactions should
result in a different class.
In spin systems the cross\-over from Heisenberg
to Ising behaviour occurs around a given temperature,
which could be very close to the
critical temperature for weak anisotropies.
There is no direct analogue of temperature in the Bak--Sneppen
model, where the critical state is now the attractor of the
dynamics, but for weak anisotropies the convergence to the
new exponents is very slow.
Nonetheless, we believe that it is the anisotropic class which should
now be regarded as the general case.

The isotropic model could still have applications in any
situation where perfect isotropy can be assumed.
The question then becomes, do any such situations exist?
In both of the model's applications we are aware of,
we think the answer is clearly `no'.
In the biological context, asymmetry between co\-evolving
species could occur for a number of reasons.
A graphic example for predator--prey relationships is known as the
``life/dinner'' principle,
where the asymmetry arises because the prey has more to
lose from a failed encounter than the predator~\cite{dawkins}.
This gets its name from an Aesop's fable,
where a dog gives up chasing a hare because it is only
running for its dinner, whereas the hare is running
for its life, hence ``life/dinner'' principle~\cite{aesop}.
Such asymmetry should result in a preferred direction
along the food chain, although the issue is somewhat
clouded here by the lack of a realistic
food web structure~\cite{webs1,webs2}.
A second application of the model has recently
been proposed for the
process whereby granular materials, such as sand, powder,
corn flakes {\em etc.}, settle under perturbations~\cite{compact}.
Here, anisotropy would be induced by gravity.

It was mentioned in the introduction that the maximally
anisotropic \mbox{$a=0$}, \mbox{$b=1$} system should be more
open to analysis than the isotropic one.
This certainly proved to be true in the
\mbox{$M\rightarrow\infty$} limit studied
in Sec.~\ref{sec:multi},
where exact solutions were found for all values
of~$\lambda$ rather than just
\mbox{$\lambda=\frac{1}{2}$}\,, as in the isotropic case.
The maximally anisotropic model
may also prove to be more tractable in the
original \mbox{$M=1$} framework.
This claim is not unreasonable and has many precedents.
For instance, the Zaitsev model is an extremal dynamical system
with similar rules to the Bak--Sneppen model, except that a random value
is subtracted from the global maximum and redistributed equally to
its nearest neighbours.
Stipulating that this value is instead only distributed in one
direction gives rise to an anisotropic variant
which can be solved exactly, including explicit expressions
for the critical exponents~\cite{Zaitsev}.
Another example is provided by the abelian sandpile
model, where a version in which the sand only
topples in one direction was solved before exact results
for the isotropic case were found~\cite{dhar1,dhar2}.

One area that we have not investigated is what
happens when anisotropy is introduced
to lattices with two or more dimensions.
This opens up the possibility of having
isotropic interactions parallel to one axis but
anisotropic interactions parallel to another,
the number of permutations between the axes
increasing with the dimensionality.
Based on the analogy with spin systems, we would expect
that there would still be just two universality classes for
each dimension, one for the fully isotropic case and
one for any non-zero anisotropy.
The critical exponents for both classes should converge
when the upper critical dimension is reached, beyond
which the introduction of anisotropy will make no difference.
Work is in progress to find at what dimension this convergence occurs,
to help confirm or deny recent claims that the upper critical
dimension for the Bak--Sneppen model is~8~\cite{ucd,in_prog}.

Just prior to publication we became aware of a modified
Bak--Sneppen model by Vendruscolo {\em et al.} which introduces
a preferred direction by a very different mechanism~\cite{corr_ev}.
Nonetheless the critical exponents for their model appear to
match those found for our one, which strengthens the case for the
universality of the anisotropic class.

%%
%%  REFERENCES
%%

\end{multicols}


\begin{references}
\bibitem{ctm} Note that this is the continuum {\em approximation}.
The continuum {\em limit}, achieved by sending the lattice spacing
to zero in a suitable manner, is not instructive here since it
introduces troublesome delta functions into the boundary conditions.
\end{references}
\end{document}